# Nonconvex $L_{1/2}$- Regularized Nonlocal Self-similarity Denoiser for Compressive Sensing based CT Reconstruction

Yunyi Li[a,*], Yiqiu Jiang[b], Hengmin Zhang[c], Jianxun Liu[a], Xiangling Ding[a], Guan Gui[d,*]

[a] School of Computer Science and Engineering, Hunan University of Science and Technology
[b] Department of Sports Medicine and Joint Surgery, Nanjing First Hospital, Nanjing Medical University
[c] Department of Computer and Information Science, University of Macau
[d] College of Telecommunications and Information Engineering, Nanjing University of Posts and Telecommunications
* Corresponding Author: yunyili@hnust.edu.cn, guiguan@njupt.edu.cn

**Abstract**: Compressive sensing (CS) based computed tomography (CT) image reconstruction aims at reducing the radiation risk through sparse-view projection data. It is usually challenging to achieve satisfying image quality from incomplete projections. Recently, the nonconvex $L_{1/2}$-norm has achieved promising performance in sparse recovery, while the applications on imaging are unsatisfactory due to its nonconvexity. In this paper, we develop a $L_{1/2}$-regularized nonlocal self-similarity (NSS) denoiser for CT reconstruction problem, which integrates low-rank approximation with group sparse coding (GSC) framework. Concretely, we first split the CT reconstruction problem into two subproblems, and then improve the CT image quality furtherly using our $L_{1/2}$-regularized NSS denoiser. Instead of optimizing the nonconvex problem under the perspective of GSC, we particularly reconstruct CT image via low-rank minimization based on two simple yet essential schemes, which build the equivalent relationship between GSC based denoiser and low-rank minimization. Furtherly, the weighted singular value thresholding (WSVT) operator is utilized to optimize the resulting nonconvex $L_{1/2}$ minimization problem. Following this, our proposed denoiser is integrated with the CT reconstruction problem by alternating direction method of multipliers (ADMM) framework. Extensive experimental results on typical clinical CT images have demonstrated that our approach can further achieve better performance than popular approaches.

**Index terms**: Compressive sensing, CT reconstruction, $L_{1/2}$, nonlocal self-similarity, group sparse coding, low-rank, ADMM.

## I. Introduction

Compressive sensing (CS) image reconstruction technique has been a powerful tool in image inverse problem, which indicates that a image can be exactly reconstructed with a high probability from insufficient under-sampling observations if the image can be sparse in some domain [1]. Over the past decade, CS theory has been widely employed in communications system [2], radar imaging[3], snapshot compressive imaging [4], remoting sensing [5] and medical imaging [6][7], etc.

As a popular technique in engineering, computed tomography (CT) has been widely applied for clinical intervention and diagnosis [8]. However, it is often harmful for human body to explore to X-ray excessively, which often causes cancerous and genetic disease [9]. Thus, it is significant to reduce the radiation dose in clinical practice, such as reduce the X-ray flux and decrease the total number of sampling views. As a result, the noisy measurements and insufficient projected data would be inevitably generated in the process of sampling and imaging. It is challenging to reconstruct high-quality CT images from degraded and insufficient projection data by traditional approaches, such as simultaneous algebraic reconstruction (SART) technique and filtered back projection (FBP) algorithm [10]. Thus, to reconstruct high-quality CT image from contaminated and under-sampled sparse-view data has attracted many attentions in CT reconstruction problem [8][11][12].

Image prior model would play a vital and effective role to restrict the solution space for CS based CT reconstruction problem. Sparsity is an intrinsic and popular prior information of signal and image, which shows that signal and image can be sparsely represented by a few atoms in a proper dictionary. The last decade has witnessed a successful application of sparsity-exploiting based methods to CS image reconstruction. Among them, the total-variation (TV) regularization [13] assumes that the image and signal is piecewise smooth, usually resulting in patchy effects. The dictionary learning based methods [14] can exploit more sparsity by learning a overcomplete dictionary. Moreover, the discrete cosine transform (DCT) [15], tight frame[16] and wavelet are also some popular sparsity models[17]. These sparsity-based CS methods have been applied to sparse-view CT image reconstruction [11][12][18][19][20]. However, the sparsity-based methods ignore the structure texture of image, often resulting in unsatisfactory CT reconstruction results. The nonlocal self-similarity (NSS) is another alternative prior for CS image reconstruction problem, recent studies have shown that the NSS prior can achieve promising performance, such as non-local mean filter [21], BM3D [22], non-locally centralized sparse representation (NCSR) [23], group sparse representation (GSR) [24][25] and nonlocal low-rank regularization [26]. Moreover, it is also important to employ a proper function for the regularization of image prior. Due to the convexity, the popular $L_1$-norm has been popularly applied to sparse recovery problem. However, the biased solution usually causes. To address this issue, some nonconvex relaxations of $L_p(0 < p < 1)$-norm, SCAD[27], and MCP[28] have been exploited widely.

Recently, image denoiser model has witnessed its promising results for many low-level image processing tasks. Actually, image denoiser is also an expert which can capsulate the statistic characteristics of images. In [29], Venkatakrishnan et al. proposed a plug-and-play (PnP) framework for imaging inverse problem, which leverages the advanced image denoising model to imaging inverse problem via the alternating direction method of multipliers (ADMM) algorithm. In [30], a nonlinear PnP framework is proposed for image reconstruction problem. In [31], a novel primal-dual PnP framework is developed based on splitting strategy. In [32], a iterative denoising and backward projections (IDBP) framework is proposed for image restoration problem. There is no denying that these PnP based imaging framework can achieve outstanding results, however, these mentioned approaches only focus on designing optimization framework while ignoring popular structural image prior, e.g., NSS. In [33], a novel denoising model is proposed for image restoration problem, which plugs a nonconvex low-rank denoiser into the ADMM framework combing with NSS. However, it is only for natural image reconstruction problem. Thus, it is also significant to apply promising denoiser model to medical imaging problem, e.g., sparse-view based CT reconstruction.

Bearing the above concern in mind, in this paper, we develop a nonconvex $L_{1/2}$-constraint regularized image reconstruction framework for CT image reconstruction, which integrates NSS as prior knowledge. The main contributions of this paper are summarized as follow: (1) We propose a novel nonconvex regularization approach for CS CT reconstruction based on traditional SART technology via the alternating minimization strategy, and develop a fast yet effective CT reconstruction algorithm. (2) We propose a $L_{1/2}$-regularized denoiser model for CT reconstruction by integrating $L_{1/2}$-norm with NSS prior, which extend the significant application of $L_{1/2}$ regularization into CT reconstruction problem. (3) We develop two essential schemes of equivalent conversion between group sparse coding model and low rank minimization based one for our denoiser, then our sparse-view CT reconstruction problem can be solved by low-rank approximation approach via ADMM framework. The flowchart of our proposed CT reconstruction approach is shown in Fig. 1.

The reminder of this paper is organized as: Section II first introduces some preliminary knowledge for our work, including CT reconstruction model, group sparse coding and $L_{1/2}$-regularization. Then, our proposed nonconvex $L_{1/2}$-regularized NSS denoiser based CT approach is presented in section III. Section IV presents the extensive experimental results on three clinical CT images and Section V concludes our work.

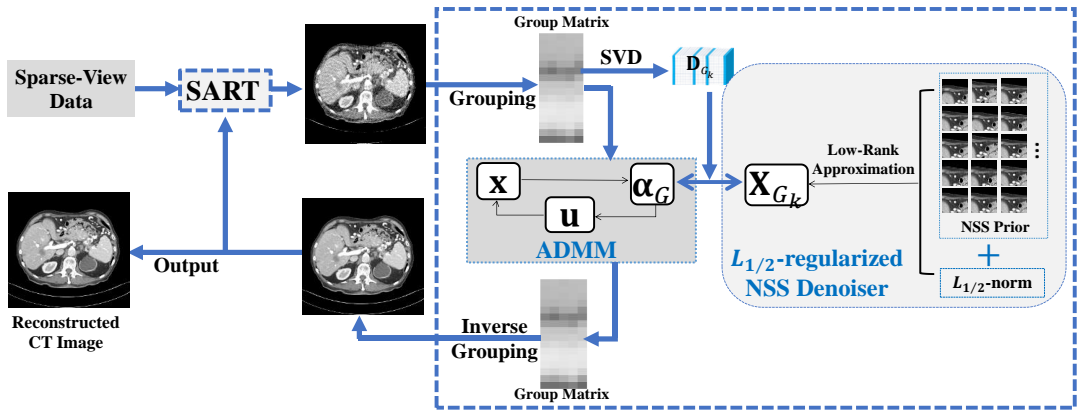

Fig. 1 The flowchart of our proposed CT reconstruction approach via nonconvex $L_{1/2}$- regularized nonlocal self-similarity denoiser

## II. Preliminary Knowledge

## 2.1. Sparse-view based CT Reconstruction Model

Typically, CT imaging system can be expressed as

$$\mathbf{b} = \mathbf{Ax} \tag{1}$$

where $\mathbf{b} \in \mathbb{R}^M$ denotes the sparse-view projection data, $\mathbf{A} \in \mathbb{R}^{M \times N}$ is the system measurement matrix ($M$ and $N$ are the total number of projection data and image pixels, respectively) and $\mathbf{x} \in \mathbb{R}^N$ presents the latent CT image. Due to the ill-posedness, the latent high-quality CT image can't reconstruction from $\mathbf{b}$ directly by simple inverse operation. According to sparse representation and CS theory, the approximate sparse solution $\hat{\mathbf{x}}$ can be obtained by minimizing the following problem:

$$\hat{\mathbf{x}} = \arg\min_{\mathbf{x}} \frac{1}{2} \|\mathbf{b} - \mathbf{Ax}\|_2^2 + \lambda \mathfrak{R}(\mathbf{x}) \tag{2}$$

where the regularization term $\mathfrak{R}(\mathbf{x})$ can provide necessary prior knowledge to restrict the solution space and $\|\mathbf{b} - \mathbf{Ax}\|_2^2$ denotes the data-fitting term to guarantee the obtained solution accords with the degradation model.

To efficiently resolve the problem (2), in this paper, we first introduce a useful auxiliary variable $\mathbf{v}$ and add the constraint $\mathbf{v} = \mathbf{x}$, then Eq. (2) can be expressed as the following constraint problem:

$$\hat{\mathbf{x}} = \arg\min_{\mathbf{x}} \frac{1}{2} \|\mathbf{b} - \mathbf{Av}\|_2^2 + \lambda \mathfrak{R}(\mathbf{x}) \quad \text{s.t.} \quad \mathbf{v} = \mathbf{x} \tag{3}$$

Accordingly, the alternating minimization approach is utilized in (3) to achieve the following sub-problems about $\mathbf{v}$ and $\mathbf{x}$, respectively:

$$\hat{\mathbf{v}} = \arg\min_{\mathbf{v}} \frac{1}{2} \|\mathbf{b} - \mathbf{Av}\|_2^2 + \lambda_1 \|\mathbf{v} - \mathbf{x}\|_2^2 \tag{4}$$

and

$$\hat{\mathbf{x}} = \arg\min_{\mathbf{x}} \frac{1}{2} \|\mathbf{v} - \mathbf{x}\|_2^2 + \lambda_2 \mathfrak{R}(\mathbf{x}) \tag{5}$$

in which, $\lambda_1$ and $\lambda_2$ are two regularization parameters, which can make a balance between the first and the second term. Accordingly, the CT reconstruction problem (2) can be split into (4) and (5). It should be noted that the problem (4) and (5) only depend on the system measurement (degradation) model and prior knowledge, respectively.

The problem (4) only depends on the system measurement model, accordingly, a very useful SART algorithm is usually employed to solve (4),

$$\mathbf{x}_j^{(p+1)} = \mathbf{x}_j^{(p)} + \frac{\theta}{A_{+,j}} \sum_{i=1}^{M} \frac{A_{i,j}}{A_{i,+}} \left( \mathbf{b}_i - \mathbf{b}_i(\mathbf{x}^{(p)}) \right) \tag{6}$$

where $\theta$ denotes the parameter, $A_{i,+} = \sum_{j=1}^{N} A_{i,j}$, $A_{+,j} = \sum_{i=1}^{M} A_{i,j}$, $\mathbf{b}(\mathbf{X}) = \mathbf{Ax}$, $(\cdot)_i$ represents the $i$-row of the matrix, $i = 1,2,\cdots,M$ and $j = 1,2,\cdots,N$.

The latent image $\mathbf{x}_j^{(p+1)}$ can be achieved by iteratively updating (6), however, the image quality will be very poor with severe artifacts due to the sparse-view measurement system, which will heavily compromise clinical value of CT image. Then, it is necessary to further improve the reconstruction quality by using image prior model with $\mathfrak{R}(\mathbf{x})$.

## 2.2. Group Sparse Coding

Recent studies have shown that the group sparse coding (GSC) framework can achieve promising results in image restoration tasks. For every image $\mathbf{x} \in \mathbb{R}^N$, which can be divided into $n$ overlapped patches $\mathbf{X}_k, i = 1,2,3,\cdots,n$ with size of $\sqrt{B_s} \times \sqrt{B_s}$, then we can construct $n$ image group $\mathbf{X}_{G_k} \in \mathbb{R}^{B_s \times c}, k = 1,2,\cdots,n$ by selecting $c$ most similar patches using K-Nearest Neighbour (KNN) approach. Then the group matrix can be sparsely represented by

$$\mathbf{X}_{G_k} = \mathbf{D}_{G_k} \boldsymbol{\alpha}_{G_k} \tag{7}$$

where $\mathbf{D}_{G_k} = [\boldsymbol{d}_{G_k,1}, \boldsymbol{d}_{G_k,2}, \cdots, \boldsymbol{d}_{G_k,m}] \in \mathbb{R}^{(B_s \times c) \times m}$ is a 3D dictionary, which can be adaptively learned from $\mathbf{X}_{G_k}$ directly [24], $\boldsymbol{\alpha}_{G_k} = [\alpha_{G_k,1}, \alpha_{G_k,2}, \cdots, \alpha_{G_k,m}] \in \mathbb{R}^{m \times 1}$ is the sparse coefficient vector. For simplicity, the whole image $\mathbf{x}$ can be sparsely represented from the point of group matrix, that is,

$$\mathbf{x} = \mathbf{D}_G \circ \boldsymbol{\alpha}_G \tag{8}$$

where the dictionary $\mathbf{D}_G$ and coefficient vector $\boldsymbol{\alpha}_G$ were generated by concatenating all $\mathbf{D}_{G_k}$ and $\boldsymbol{\alpha}_{G_k}$, respectively.

### 2.3. Low-rank Minimization

The method of low-rank minimization can recover the latent low-rank matrix from its degraded observation with a variety of successful applications, including signal processing [34][35], telecommunications system[36][37][38][39], medical image processing [40][41][42] and machine learning [43]. To be concrete, for a matrix with low-rank property, e.g., the image group matrix $\mathbf{X}_{G_k}$, we can achieve its approximation from the observation $\mathbf{Y}_{G_k}$ by following rank-regularized minimization problem [44]

$$\min_{\mathbf{X}_{G_k}} \frac{1}{2} \|\mathbf{Y}_{G_k} - \mathbf{X}_{G_k}\|_F^2 + \lambda rank(\mathbf{X}_{G_k}) \tag{9}$$

where $rank(\cdot)$ denotes the $L_0$-norm of the singular value vector. To solve this challenging NP-hard optimization problem, the $rank(\mathbf{X}_{G_k})$ is usually relaxed by convex nuclear norm [45].

### 2.3. Nonconvex $L_{1/2}$-regularization

Generally, the $L_1$-norm penalties and nuclear norm can achieve reliable sparse solution and the low-rank matrix with high probability. However, the bias estimates by both of them is an inevitable problem. As a representative nonconvex relaxation function, recent studies have shown that nonconvex $L_{1/2}$-norm can obtain more

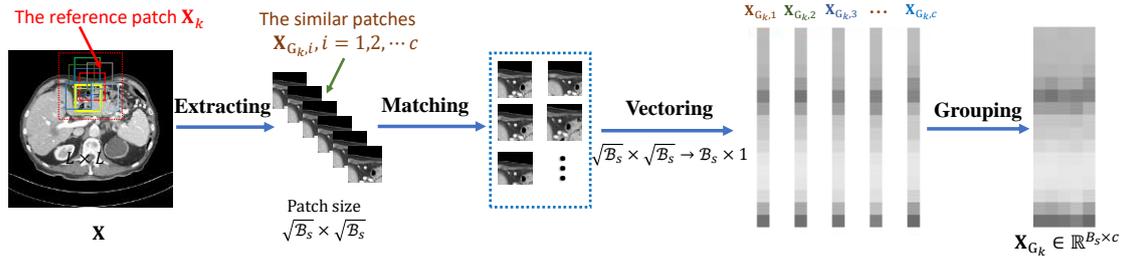

Fig.2 The construction of group matrix $\mathbf{X}_{G_k}$

accurate sparse solution than $L_1$-norm, which has demonstrated promising performance in sparse recovery, image restoration tasks and machine learning. Considering the following sparse signal recovery problem about $\mathbf{x} \in \mathbb{R}^N$,

$$\hat{\mathbf{x}} = \arg\min_{\mathbf{x}} \frac{1}{2} \|\mathbf{y} - \mathbf{x}\|_2^2 + \lambda \|\mathbf{x}\|_{1/2}^{1/2} \tag{10}$$

where $\lambda$ denotes the regularization parameter, and $\|\mathbf{x}\|_{1/2}^{1/2} = \sum_{i=1}^{N} \sqrt{|x_i|}$ denotes the $L_{1/2}$-norm [46][47][48][49]. The iterative half thresholding algorithm can be defined

as
$$\mathbf{x}^{(t+1)} = \mathbf{H}_{\lambda,1/2}\left(\left(\theta\mathbf{x}^{(t)}\right)\right) \tag{11}$$
where the operator $\theta(\cdot): \mathbb{R}^N \to \mathbb{R}^N$ is an affine transform, $\theta(\mathbf{x}^{(t)}) = \mathbf{x}^{(t)} - \tau\mathbf{\Phi}^T(\mathbf{\Phi}\mathbf{x}^{(t)} - \mathbf{y})$ and $\tau > 0$ is the step size. In which, the thresholding operator $\mathbf{H}_{\lambda,1/2}(\cdot)$ is defined as
$$\mathbf{H}_{\lambda,1/2}(\mathbf{x}) = [h(x_1), h(x_2), \cdots, h(x_N)]^T \tag{12}$$
and each function $h(\cdot)$ is defined by:
$$h_{\lambda,1/2}(x_i) = \begin{cases} \varphi_{\lambda,1/2}(x_i), & |x_i| > T \\ 0, & otherwise \end{cases} \tag{13}$$
where $\varphi_{\lambda,1/2}(x_i) = \frac{2}{3}x_i\left\{1 + \cos\left[\frac{2\pi}{3} - \frac{2}{3}\cos^{-1}\left(\frac{\lambda}{8}\left(\frac{|x_i|}{3}\right)^{-\frac{2}{3}}\right)\right]\right\}$, $T = \frac{3\sqrt[3]{2}}{4}\lambda^{2/3}$ is the thresholding value.

## III. Nonconvex $L_{1/2}$-regularized NSS Denoiser based CT Reconstruction Model

### 3.1. Nonconvex $L_{1/2}$-Regularized Nonlocal Self-similarity Denoiser

To improve the CT reconstruction using nonconvex $L_{1/2}$-regularization and the NSS prior for optimization problem (5). Firstly, we need to build the nonconvex $L_{1/2}$-regularized NSS denoiser. According to the GSC framework, the corresponding low-rank denoising model can be expressed as
$$\hat{\boldsymbol{\alpha}}_{G_k} = \arg\min_{\boldsymbol{\alpha}_{G_k}} \frac{1}{2}\|\mathbf{Y}_{G_k} - \mathbf{D}_{G_k}\boldsymbol{\alpha}_{G_k}\|_F^2 + \lambda\|\boldsymbol{\alpha}_{G_k}\|_0 \tag{14}$$
where $\mathbf{Y}_{G_k}$ denotes the corresponding observation group matrix. Obviously, it is difficult to solve the $L_0$-based NP hard problem. To effectively and efficiently address this denoising problem, two essential schemes is developed to build the theory connection for the further relaxation and conversion, namely *Scheme 1* and *Scheme 2*.

*Scheme 1.*

We first bridge the important gap between GSC based denoiser model and low-rank minimization based denoising problem. Then the low-rank minimization problem is relaxed into nonconvex low-rank approximation problem.

---

**Remark 3.1** [33]. *For group data matrix $\mathbf{X}_{G_k} = \mathbf{D}_{G_k}\boldsymbol{\alpha}_{G_k}$, and its singular value decomposition (SVD) is $\mathbf{X}_{G_k} = \sum_{i=1}^m \sigma_{\mathbf{X}_{G_k},i}\boldsymbol{u}_{\mathbf{X}_{G_k},i}\boldsymbol{v}_{\mathbf{X}_{G_k},i}^T$, we have the following relationship*
$$\|\boldsymbol{\alpha}_{G_k}\|_0 = rank\left(\sum_{i=1}^m \sigma_{\mathbf{X}_{G_k},i}\mathbf{d}_{G_k,i}\right) = rank(\mathbf{D}_{G_k}\boldsymbol{\alpha}_{G_k}) = rank(\mathbf{X}_{G_k}) \tag{15}$$
*in which, the vector $\boldsymbol{\alpha}_{G_k} = [\alpha_{G_k,1}, \alpha_{G_k,2}, \cdots, \alpha_{G_k,m}] \in \mathbb{R}^{m\times 1}$, the dictionary $\mathbf{D}_{G_k} = [\boldsymbol{d}_{G_k,1}, \cdots, \boldsymbol{d}_{G_k,m}] \in \mathbb{R}^{(B_s\times c)\times m}$, and the $rank(\mathbf{X}_{G_k})$ denotes the singular value number of matrix $\mathbf{X}_{G_k}$.*

---

According to **Remark 3.1**, if we let $\mathbf{X}_{G_k} = \mathbf{D}_{G_k}\boldsymbol{\alpha}_{G_k}$, then the group-based CT denoising problem (14) is equivalent to the following low-rank minimization problem

$$\widehat{\mathbf{X}}_{G_k} = \arg\min_{\mathbf{X}_{G_k}} \frac{1}{2} \|\mathbf{Y}_{G_k} - \mathbf{X}_{G_k}\|_F^2 + \lambda rank(\mathbf{X}_{G_k}) \tag{16}$$

Usually, the minimization problem (16) is NP-hard. To estimate the approximation of low-rank matrix $\widehat{\mathbf{X}}_{G_k}$ with barely bias, in this paper, we utilize the weight nonconvex function as the relaxation function, that is

$$\widehat{\mathbf{X}}_{G_k} = \arg\min_{\mathbf{X}_{G_k}} \frac{1}{2} \|\mathbf{Y}_{G_k} - \mathbf{X}_{G_k}\|_F^2 + \lambda \hbar_{w_k}\left(\sigma(\mathbf{X}_{G_k})\right) \tag{17}$$

where $\hbar_{w_k}\left(\sigma(\mathbf{X}_{G_k})\right) = \sum_{i=1}^{r} w_{k,i} \hbar(\sigma_{k,i})$, $r = min(B_s, c)$, $\mathbf{w}_k = (w_{k,1}, w_{k,2}, \cdots, w_{k,r})$ denotes the weight vector with $w_{k,1} \leq w_{k,2} \leq \cdots \leq w_{k,r}$, $\sigma(\mathbf{x}_{G_k}) = (\sigma_{k,1}, \sigma_{k,2}, \cdots, \sigma_{k,r})$ denotes the singular value vector of group matrix $\mathbf{x}_{G_k}$ with $\sigma_{k,1} \leq \sigma_{k,2} \leq \cdots \leq \sigma_{k,r}$, $\hbar(\cdot): \mathbb{R}^+ \to \mathbb{R}^+$ denotes the square root function, e.g., $\hbar(x_i) = |x_i|^{1/2}$, thus $\hbar(\cdot)$ is proper and lower semi-continuous is nondecreasing on $[0, +\infty)$.

*Scheme 2.*
　　We first relax the NP hard problem (14) with nonconvex optimization problem. Then we build an equivalent connection between the nonconvex GSC denoising problem and the nonconvex low-rank matrix approximation problem using an adaptive group dictionary.

> **Remark 3.2** [24] *For group matrix $\mathbf{X}_{G_k}$, if its singular value decomposition (SVD) is defined as*
> $$\mathbf{X}_{G_k} = \mathbf{U}_{\mathbf{X}_{G_k}} \mathbf{\Sigma}_{\mathbf{X}_{G_k}} \mathbf{V}_{\mathbf{X}_{G_k}}^T = \sum_{i=1}^{m} \sigma_{\mathbf{X}_{G_k},i} \mathbf{u}_{\mathbf{X}_{G_k},i} \mathbf{v}_{\mathbf{X}_{G_k},i}^T \tag{18}$$
> *the atoms $\mathbf{d}_{G_k,i}$, $i = 1,2,\cdots,m$ can be defined by*
> $$\mathbf{d}_{G_k,i} = \mathbf{u}_{\mathbf{X}_{G_k},i} \mathbf{v}_{\mathbf{X}_{G_k},i}^T, \ i = 1,2,\cdots,m. \tag{19}$$
> *Accordingly, the self-adaptive dictionary for each group can be given by*
> $$\mathbf{D}_{G_k} = [\mathbf{d}_{G_k,1}, \mathbf{d}_{G_k,2}, \cdots, \mathbf{d}_{G_k,m}]. \tag{20}$$
> *where* $\mathbf{U}_{G_k} = [\mathbf{u}_{\mathbf{X}_{G_k},1}, \mathbf{u}_{\mathbf{X}_{G_k},2}, \cdots, \mathbf{u}_{\mathbf{X}_{G_k},m}]$, $\mathbf{V}_{\mathbf{X}_{G_k}} = [\mathbf{v}_{\mathbf{X}_{G_k},1}, \mathbf{v}_{\mathbf{X}_{G_k},2}, \cdots, \mathbf{v}_{\mathbf{X}_{G_k},m}]$ *and* $\mathbf{\Sigma}_{\mathbf{X}_{G_k}} = diag([\sigma_{\mathbf{X}_{G_k},1}; \sigma_{\mathbf{X}_{G_k},2}; \cdots; \sigma_{\mathbf{X}_{G_k},m}])$, *in which* $m = min(B_s, c)$.

The problem (14) can be first relaxed into the following nonconvex group-sparse recovery problem:

$$\widehat{\boldsymbol{\alpha}}_{G_k} = \arg\min_{\boldsymbol{\alpha}_{G_k}} \frac{1}{2} \|\mathbf{Y}_{G_k} - \mathbf{D}_{G_k} \boldsymbol{\alpha}_{G_k}\|_F^2 + \lambda \hbar_{w_k}(\boldsymbol{\alpha}_{G_k}) \tag{21}$$

where $\hbar_{w_k}(\boldsymbol{\alpha}_{G_k}) = \sum_{i=1}^{m} w_{k,i} \hbar(\alpha_{k,i})$, $\mathbf{w}_k = (w_{k,1}, w_{k,2}, \cdots, w_{k,r})$ denotes the weight vector, $\hbar(\cdot): \mathbb{R}^+ \to \mathbb{R}^+$ denotes the square root function, e.g., $\hbar(x_i) = |x_i|^{1/2}$, thus $\hbar(\cdot)$ is proper and lower semi-continuous is nondecreasing on $[0, +\infty)$. The **Remark 3.2** shows that the singular value of adaptive dictionary $\mathbf{D}_{G_k}$ can be regarded as the sparse coefficients of $\mathbf{Y}_{G_k}$ by (13), then we have the following equation of (13), i.e.,

$$\widehat{\boldsymbol{\alpha}}_{G_k} = \arg\min_{\boldsymbol{\alpha}_{G_k}} \frac{1}{2} \|\mathbf{Y}_{G_k} - \mathbf{D}_{G_k} \boldsymbol{\alpha}_{G_k}\|_F^2 + \lambda \hbar_{w_k}(\mathbf{D}_{G_k} \boldsymbol{\alpha}_{G_k}) \tag{22}$$

After substituting $\mathbf{D}_{G_k} \boldsymbol{\alpha}_{G_k}$ with $\mathbf{X}_{G_k}$, the optimization problem (14) can be equivalently converted into problem (17), for convenience, we rewrite the problem (17) as here,

$$\widehat{\mathbf{X}}_{G_k} = \arg\min_{\mathbf{X}_{G_k}} \frac{1}{2}\|\mathbf{Y}_{G_k} - \mathbf{X}_{G_k}\|_F^2 + \lambda \hbar_{w_k}\left(\sigma(\mathbf{X}_{G_k})\right) \quad (23)$$

It is difficult to minimize the problem (17) or (23) due to the non-convexity and non-smoothness of $\hbar_{w_k}(\cdot)$, we first give the **Theorem 3.1**:

---

**Theorem 3.1**[50] *If matrix $\mathbf{Y} \in \mathbb{R}^{M \times N}$, $\hbar_w(\sigma(\mathbf{X})) = \sum_{i=1}^{r} w_i \rho(\sigma_i)$, the $\hbar(\cdot): \mathbb{R}^+ \to \mathbb{R}^+$ denotes the proper and lower semi-continuous function, and is nondecreasing on $[0, +\infty)$, the weighting vector $\mathbf{w} = [w_1, w_2, \cdots, w_s]$ with $0 \leq w_1 \leq w_2 \leq \cdots \leq w_s$, $(s = \min(M, N))$, then for any parameter $\lambda > 0$, the optimal solution of the following optimization problem*

$$\min \frac{1}{2}\|\mathbf{Y} - \mathbf{X}\|_F^2 + \lambda \hbar_w(\sigma(\mathbf{X})) \quad (24)$$

*can be achieved by the following weighted singular value thresholding (WSVT) operator,*

$$\widehat{\mathbf{X}} = \mathbf{u} Diag\left(\text{Prox}_{w,\rho}^{\delta}(\boldsymbol{\delta}(\mathbf{Y}))\right)\mathbf{v}^{\mathrm{T}} \quad (25)$$

*where the element-wise proximal operator $\text{Prox}_{w,\rho}^{\delta}(\boldsymbol{\delta}(\mathbf{Y}))$ is defined by*

$$\text{Prox}_{w_i,\rho}^{\delta_i}(\delta_i(\mathbf{Y})) = \arg\min_{\delta_i > 0} \frac{1}{2}\left(\sigma_i(\mathbf{X}_{G_k}) - \delta_i(\mathbf{Y}_{G_k}^{(t)})\right)^2 + \lambda w_{k,i}^{(t)} \rho\left(\sigma_i(\mathbf{X}_{G_k})\right) \quad (26)$$

---

According to **Theorem 3.1** the nonconvex $L_{1/2}$-Constraint low-rank minimization problem (13) earns the following Global optimal solution,

$$\mathbf{X}_{G_k}^{(t+1)} = \mathbf{u}_{G_k} Diag\left(\text{Prox}_{w_k,\rho}^{\delta}\left(\boldsymbol{\delta}(\mathbf{Y}_{G_k}^{(t)})\right)\right)\mathbf{v}_{G_k}^{\mathrm{T}} \quad (27)$$

in which, $\mathbf{u}_{G_k}$ and $\mathbf{v}_{G_k}^{\mathrm{T}}$ is obtain by the SVD of $\mathbf{X}_{G_k}$, that is $\mathbf{X}_{G_k} = \mathbf{u}_{G_k} \boldsymbol{\Sigma}_{G_k} \mathbf{v}_{G_k}^{\mathrm{T}}$, and the proximal operator $\text{Prox}_{w_k,\rho}^{\delta}\left(\boldsymbol{\delta}(\mathbf{Y}_{G_k}^{(t)})\right)$ is defined as,

$$\text{Prox}_{w_i,\rho}^{\delta_i}\left(\delta_i(\mathbf{Y}_{G_k}^{(t)})\right) = \arg\min_{\delta_i > 0} \frac{1}{2}\left(\sigma_i(\mathbf{X}_{G_k}) - \delta_i(\mathbf{Y}_{G_k}^{(t)})\right)^2 + \lambda w_{k,i}^{(t)} \rho\left(\sigma_i(\mathbf{X}_{G_k})\right) \quad (28)$$

where $\rho(\sigma_i) = (\sigma_i)^{1/2}$, then the sparse solution of (18) can be achieved by iterative half thresholding algorithm [46].

### 3.2. Integrating Nonconvex $L_{1/2}$-regularized Denoiser with CT reconstruction

As mentioned in section 2.1, the reconstructed CT image by SART would suffer from poor quality. To improve the CT reconstruction performance, in this subsection, we will introduce how to utilized our proposed nonconvex $L_{1/2}$-constraint group sparsity denoiser to elevate the image quality by solving the minimization problem (5).

We first introduce a constraint condition, e.g., $\mathbf{x} = \mathbf{D}_G \circ \boldsymbol{\alpha}_G$, then the minimization problem (5) can be rewritten as,

$$\widehat{\mathbf{x}}, \widehat{\boldsymbol{\alpha}}_G = \arg\min_{\mathbf{x},\boldsymbol{\alpha}_G} \frac{1}{2}\|\mathbf{v} - \mathbf{x}\|_2^2 + \lambda_2 \Re(\boldsymbol{\alpha}_G), \quad s.t. \quad \mathbf{x} = \mathbf{D}_G \circ \boldsymbol{\alpha}_G \quad (29)$$

According to the alternative optimization strategy, Eq. (29) can be transformed into following three iterative steps:

$$\begin{cases} \mathbf{x}^{(t+1)} = \arg\min_{\mathbf{z}} \frac{1}{2}\|\mathbf{v} - \mathbf{x}\|_2^2 + \frac{\mu}{2}\left\|\mathbf{x} - \mathbf{D}_G \circ \boldsymbol{\alpha}_G^{(t)} - \mathbf{u}^{(t)}\right\|_2^2 \\ \boldsymbol{\alpha}_G^{(t+1)} = \arg\min_{\boldsymbol{\alpha}_G} \frac{\mu}{2}\left\|\mathbf{x}^{(t+1)} - \mathbf{D}_G \circ \boldsymbol{\alpha}_G - \mathbf{u}^{(t)}\right\|_2^2 + \lambda_2 \Re(\boldsymbol{\alpha}_G) \\ \mathbf{u}^{(t+1)} = \mathbf{u}^{(t)} - \left(\mathbf{x}^{(t+1)} - \mathbf{D}_G \circ \boldsymbol{\alpha}_G^{(t+1)}\right) \end{cases} \quad (30)$$

The Eq. (30) includes two independent subproblems of $\mathbf{x}$ and $\boldsymbol{\alpha}_G$. Actually, the subproblem of $\mathbf{x}$ is a typical quadratic optimization problem, and the closed-form solution is

$$\mathbf{x} = (\mathbf{I} + \mu\mathbf{I})^{-1}\big(\mathbf{v} + \mu(\mathbf{D}_G \circ \boldsymbol{\alpha}_G + \mathbf{u})\big) \tag{31}$$

in which, $\mathbf{I}$ denotes the identity matrix. For problem $\boldsymbol{\alpha}$, if we replace $\mathbf{x}^{(t+1)} - \mathbf{u}^{(t)}$ with $\mathbf{r}^{(t+1)}$, e.g., $\mathbf{r}^{(t+1)} = \mathbf{x}^{(t+1)} - \mathbf{u}^{(t)}$, then the subproblem $\boldsymbol{\alpha}_G$ can be transform the following denoising problem,

$$\boldsymbol{\alpha}_G^{(t+1)} = \arg\min_{\boldsymbol{\alpha}_G} \frac{\mu}{2}\left\|\mathbf{r}^{(t+1)} - \mathbf{D}_G \circ \boldsymbol{\alpha}_G\right\|_2^2 + \lambda_2 \mathfrak{R}(\boldsymbol{\alpha}_G) \tag{32}$$

Recalling the relationship of Eq. (15) in *scheme* **1** or the relationship between Eq. (21) and (23) in *scheme* **2**, the CT reconstruction sub-problem (32) can be equal to the following low-rank minimization problem,

$$\mathbf{x}_G^{(t+1)} = \arg\min_{\mathbf{X}_G} \frac{\mu}{2}\left\|\mathbf{r}^{(t+1)} - \mathbf{X}_G\right\|_2^2 + \lambda_2 \mathfrak{R}(\mathbf{x}_G) \tag{33}$$

where $\mathbf{X}_G = \mathbf{D}_G \circ \boldsymbol{\alpha}_G$.

Considering the relationship between image $\mathbf{x}$ and its group-matrix $\mathbf{X}_{G_k}$, when $N \to \infty$ and $K \to \infty$, the following equality relationship holds with very high probability (near to 1) [24][51],

$$\frac{1}{N}\|\mathbf{r} - \mathbf{X}_G\|_2^2 = \frac{1}{K}\sum_{k=1}^{n}\left\|\mathbf{R}_{G_k} - \mathbf{X}_{G_k}\right\|_F^2 \tag{34}$$

where $K = n \times c \times B_s$, then the problem (26) can be transformed into the following minimization problem:

$$\widehat{\mathbf{X}}_G = \arg\min_{\mathbf{X}_{G_k}} \frac{1}{2}\sum_{k=1}^{m}\left\|\mathbf{X}_{G_k} - \mathbf{R}_{G_k}\right\|_F^2 + \tau \sum_{k=1}^{m} \mathfrak{R}(\mathbf{X}_{G_k}) \tag{35}$$

in which $\tau = \frac{\lambda_2 K}{\mu N}$. For each group matrix $\mathbf{X}_{G_k}$, we have

$$\widehat{\mathbf{X}}_{G_k} = \arg\min_{\mathbf{X}_{G_k}} \frac{1}{2}\left\|\mathbf{X}_{G_k} - \mathbf{R}_{G_k}\right\|_F^2 + \tau \mathfrak{R}(\mathbf{X}_{G_k}) \tag{36}$$

To improve the CT reconstruction quality, the weight nonconvex function $\hbar_{w_k}(\cdot)$ is utilized to replace the regularization term $\mathfrak{R}(\mathbf{X}_{G_k})$, then we have

$$\widehat{\mathbf{X}}_{G_k} = \arg\min_{\mathbf{X}_{G_k}} \frac{1}{2}\left\|\mathbf{X}_{G_k} - \mathbf{R}_{G_k}\right\|_F^2 + \tau \hbar_{w_k}\big(\sigma(\mathbf{X}_{G_k})\big) \tag{37}$$

Thus, recalling the proposed nonconvex $L_{1/2}$-regularized NSS denoiser in subsection 3.1, the solution of problem (37) can be achieved by

$$\mathbf{X}_{G_k}^{(t+1)} = \mathbf{U}_{R_{G_k}} \text{Prox}_\rho^\delta\big(\delta\big(\mathbf{R}_{G_k}^{(t)}\big)\big) \mathbf{V}_{R_{G_k}}^T, \quad i = 1, \cdots, \min(B_s, c) \tag{38}$$

Recalling $L_{1/2}$ into $\hbar(\sigma_{k,i})$, we have

$$\sigma_i^*(\mathbf{X}_{G_k}) = \arg\min_{\delta_i > 0} \frac{1}{2}\big(\sigma_i(\mathbf{X}_{G_k}) - \delta_i\big(\mathbf{Y}_{G_k}^{(t)}\big)\big)^2 + \xi_i\big(\sigma_i(\mathbf{X}_{G_k})\big)^{1/2} \tag{39}$$

Then recalling the iterative half thresholding algorithm of $L_{1/2}$, the closed-form solution of (39) as be expressed as

$$\sigma_i^*(\mathbf{X}_{G_k}) = \begin{cases} \frac{2}{3}\sigma_i(\mathbf{X}_{G_k})\left(1 + \cos\left(\frac{2\pi}{3} - \frac{2}{3}\varphi\left(\sigma_i(\mathbf{X}_{G_k})\right)\right)\right), & |\sigma_i(\mathbf{X}_{G_k})| > T' \\ 0, & \text{otherwise} \end{cases} \quad (40)$$

in which, the function $\varphi\left(\sigma_i(\mathbf{X}_{G_k})\right) = \cos^{-1}\left(\frac{\xi_i}{4}\left(\frac{|\sigma_i(\mathbf{X}_{G_k})|}{3}\right)^{-3/2}\right)$, $\xi_i = \left(\lambda w_{k,i}^{(t)}\right)$ and the thresholding $T' = \frac{3\sqrt[3]{2}}{4}(2\xi_i)^{2/3}$, where $\lambda$ denotes the regularization parameter, which can be set manually, and $w_{k,i}^{(t)}$ denotes the weighting, which can be updated in Eq. (17).

Thus, we can reconstruct CT image via alternatively updating (4) and (5). The whole algorithm can be summarized as the **Algorithm I**.

---

**Algorithm 1** $\hat{\mathbf{x}} = \arg\min_{\mathbf{x}} \frac{1}{2}\|\mathbf{b} - \mathbf{A}\mathbf{x}\|_2^2 + \lambda \Re(\mathbf{x})$

---
**Input:** $\mathbf{b}$, $\mathbf{A}$;
**Initialization:** $c$, $\mathcal{B}_s$, $t = 0$, $\lambda$, $\mu$, $\gamma$, $\mathbf{x}^{(0)}$, $\boldsymbol{\alpha}^{(0)}$, $\mathbf{u}^{(0)}$;
**Repeat:**
 （1）**For subproblem (4) by calling SART;**
 （2）**For subproblem (5):**
  Updating $\mathbf{x}^{(t+1)}$ by Eq. (31);
  Computing $\mathbf{r}^{(t+1)} = \mathbf{x}^{(t+1)} - \mathbf{u}^{(t)}$;
  Constructing the group matrix $\{\mathbf{R}_{G_k}^{(t+1)}\}$ using $\mathbf{r}^{(t+1)}$;
  **For each group $\mathbf{R}_{G_k}$, repeat:**
   Learn adaptive dictionary $\mathbf{D}_{G_k}$ by using Eq. (19) and Eq. (20);
   Obtain $\boldsymbol{\alpha}_{G_k}$ and $\mathbf{X}_{G_k}$ by minimizing Eq. (32);
  **end for**
  Obtain $D^{(t+1)}$ by concatenating all corresponding $\{D_{G_k}\}$;
  Updating $\alpha_G^{(t+1)}$ by concatenating all corresponding $\{\alpha_{G_k}\}$;
  Updating $u^{(t+1)}$;
  Comuting $\mathbf{x}^{(t+1)}$ by concatenating all corresponding group matrix $\{\mathbf{X}_{G_k}\}$;
 $t = t + 1$;
**Output: Reconstructed CT image $\hat{\mathbf{x}}$**

---

## IV. Experimental Results

To evaluate the performance and effectiveness of our proposed NSS denoiser for CS CT reconstruction, this section would conduct extensive experiments using 3 typical real clinical data, include abdominal, pelvic and thoracic (https://imaging.nci.nih.gov/ncia/). The sparse-view projection data are generated by Siddon's ray-driven algorithm [52] under noiseless condition, and there are 64 projection views are distributed over 360° evenly. Some important parameters are set as follow: $\sqrt{\mathcal{B}_s} \times \sqrt{\mathcal{B}_s} = 32 \times 32$, the patch size is set to $8 \times 8$, the searching window size $L \times L$ is set to $20 \times 20$, and the internal iteration number is 5. We choose three popular method for comparison, including SART, GSR-NNM and GSR-SART [7]. To reduce the impact of parameters on other reconstruction results, we also set appropriate parameters for all algorithms for comparison. All experiments were conducted in **MATLAB R2017b**.

To evaluate the reconstruction performance of different algorithms, the PSNR, RMSE and structural similarity (SSIM) were utilized to quantitatively evaluate the reconstruction quality. Let $\mathbf{x}$ denotes the reference CT image, and $\hat{\mathbf{x}}$ is the reconstructed CT image. Then the quantization index of peak signal to noise ratio (PSNR) is defined by:

$$PSNR(\mathbf{x},\hat{\mathbf{x}}) = 10 log_{10} \frac{255 \times 255}{MSE(\mathbf{X},\hat{\mathbf{X}})} \quad (41)$$

where $MSE(\mathbf{x},\hat{\mathbf{x}}) = \frac{1}{N_1 \times N_2} \sum_{i=1}^{N_1} (\mathbf{x}_{ij} - \hat{\mathbf{x}}_{ij})^2$ denotes the mean squared error. The root mean square error (RMSE) is defined by:

$$RMSE(\mathbf{x},\hat{\mathbf{x}}) = \sqrt{\frac{1}{N_1 \times N_2} \sum_{i=1}^{N_1} (\mathbf{x}_{ij} - \hat{\mathbf{x}}_{ij})^2} \quad (42)$$

The structural similarity index metric (SSIM) [53] is obtained by:

$$SSIM(\mathbf{x},\hat{\mathbf{x}}) = \frac{(2\mu_\mathbf{x}\mu_{\hat{\mathbf{x}}}+C_1)(2\sigma_{\mathbf{x}\hat{\mathbf{x}}}+C_2)}{(\mu_\mathbf{x}^2+\mu_{\hat{\mathbf{x}}}^2+C_1)(\sigma_\mathbf{x}^2+\sigma_{\hat{\mathbf{x}}}^2+C_2)} \quad (43)$$

in which, $\mu_\mathbf{x}$ and $\mu_{\hat{\mathbf{x}}}$ denote the mean values of $\mathbf{x}$ and $\hat{\mathbf{x}}$, respectively, $C_1$ and $C_2$ are two constants, and $\sigma_\mathbf{x}^2$ and $\sigma_{\hat{\mathbf{x}}}^2$ denote the variances of $\mathbf{x}$ and $\hat{\mathbf{x}}$, respectively.

### 4.1. Abdominal Case

We first consider the abdominal clinical data. There are two important parameters in our proposed nonconvex method, which is manually setting as $\lambda = 0.000025$ and $\mu = 0.19$. Table I shows all results achieved by different algorithms. It is obviously that our proposed algorithm can achieve best quantitative results for all metric. To visualize the reconstruction results, the corresponding CT image results are shown in Fig. 2. In Fig. 2 (b), the results of SART contains serious artifacts and is useless for clinical diagnosis with its worst-case performance. In Fig. 2 (c), the convex algorithm of GSR-NNM may reconstruct CT image, but some artifacts may also exist. In Fig. 2 (d), the approach of GSR-SART may effectively reconstruct CT image, however, some details may also be over-smoothed. Fig. 2 (e) shows that our proposed nonconvex method can achieve better CT reconstruction result with more abundant details. Fig.2 (f)-(j) are corresponding magnified region of our interest of CT image (a)-(e), and Fig.2 (k)-(n) are reconstruction relative error $abs(\mathbf{x} - \hat{\mathbf{x}})$ achieved by SART, GSR-NNM, GSR-SART and our proposed algorithm. All performances strongly demonstrate the ability of our proposed nonconvex $L_{1/2}$-group constraint denoiser, which can effectively preserve more structure and reduce artifacts for CT image reconstruction. To further evaluate the difference and superiority between our algorithm and the competing approaches, Fig. 3 and Fig. 4 visualize the horizontal (128th row) and vertical (128th column) profiles of the abdominal image, respectively. In which, the profiles achieved by our proposed nonconvex algorithm are obviously closer to the reference profiles.

**Table I**. Quantitative results reconstructed by different methods for abdominal

|  | PSNR | RMSE | SSIM |
|---|---|---|---|
| SART | 32.98 dB | 0.0224 | 0.8640 |
| GSR-NNM | 45.01 dB | 0.0056 | 0.9826 |
| GSR-SART | 46.88 dB | 0.0045 | 0.9860 |
| Proposed | **47.60 dB** | **0.0042** | **0.9881** |

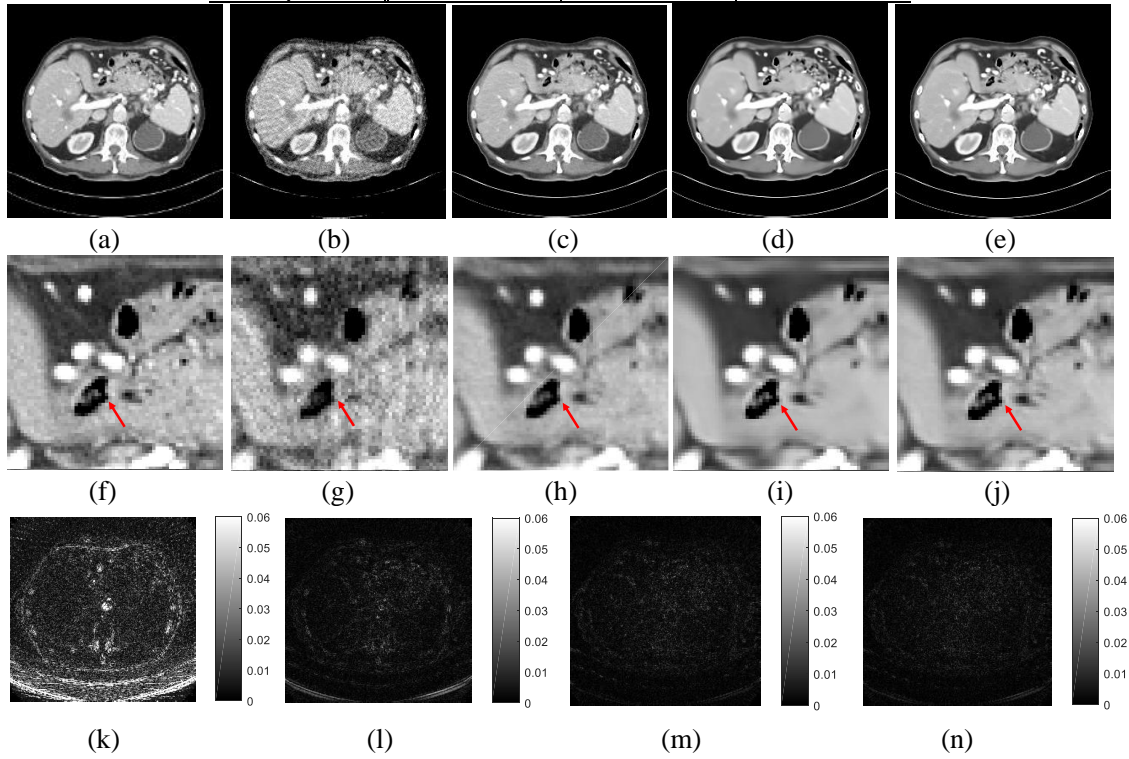

Fig. 2. Reconstructed results of abdominal images. (a) Original image, (b) Reconstructed image by SART, (c) Reconstructed image by GSR-NNM, (d) Reconstructed image by GSR-SART, (e) Reconstructed image by our proposed nonconvex algorithm; (f)-(j) The magnified region of our interest of CT image (a)-(e), (k)-(n) The relative error achieved by SART, GSR-NNM, GSR-SART and our proposed algorithm.

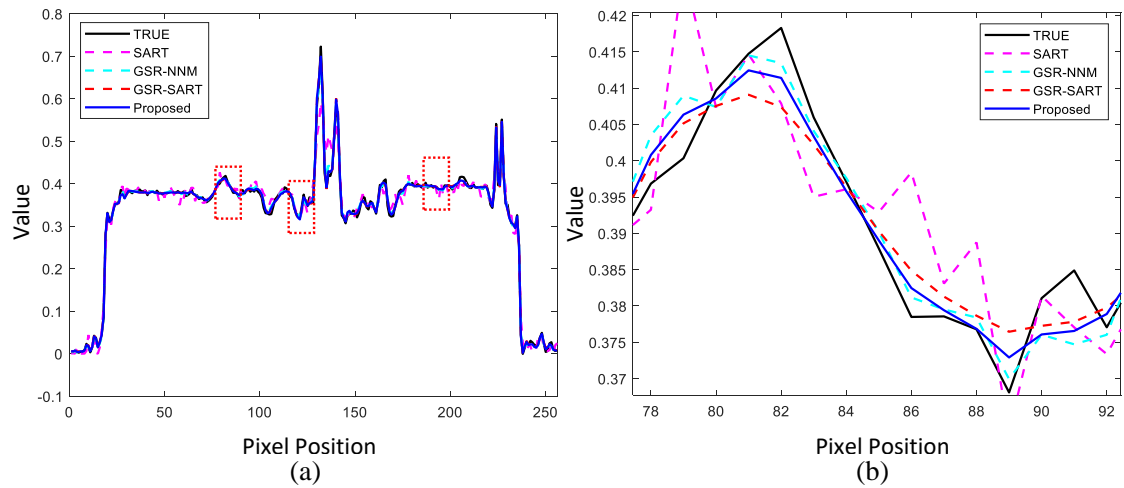

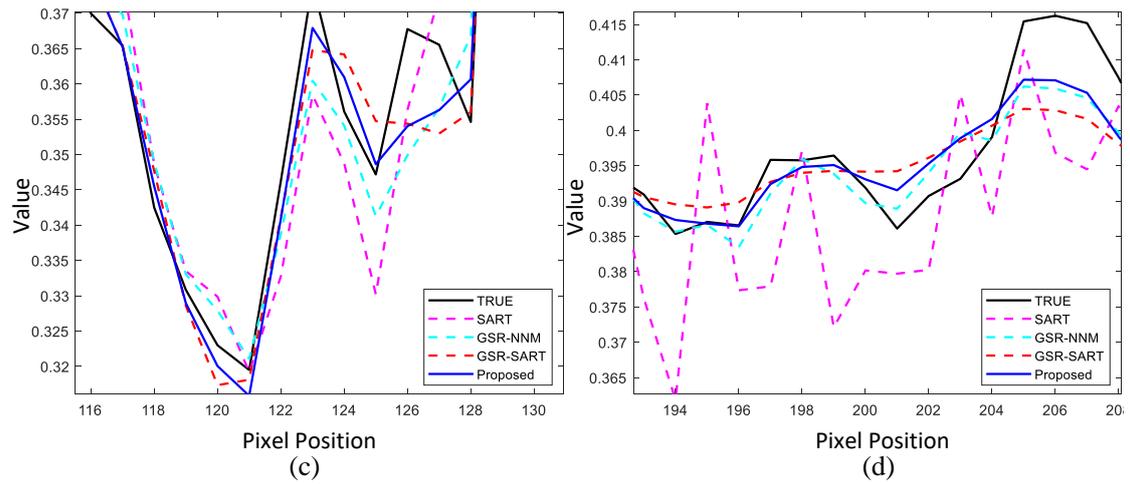

Fig.3. Horizontal profiles (128th row) of the abdominal and several generated profiles by different algorithms. (a) The overall profiles; (b)-(d) The magnified regions of our interest of CT image.

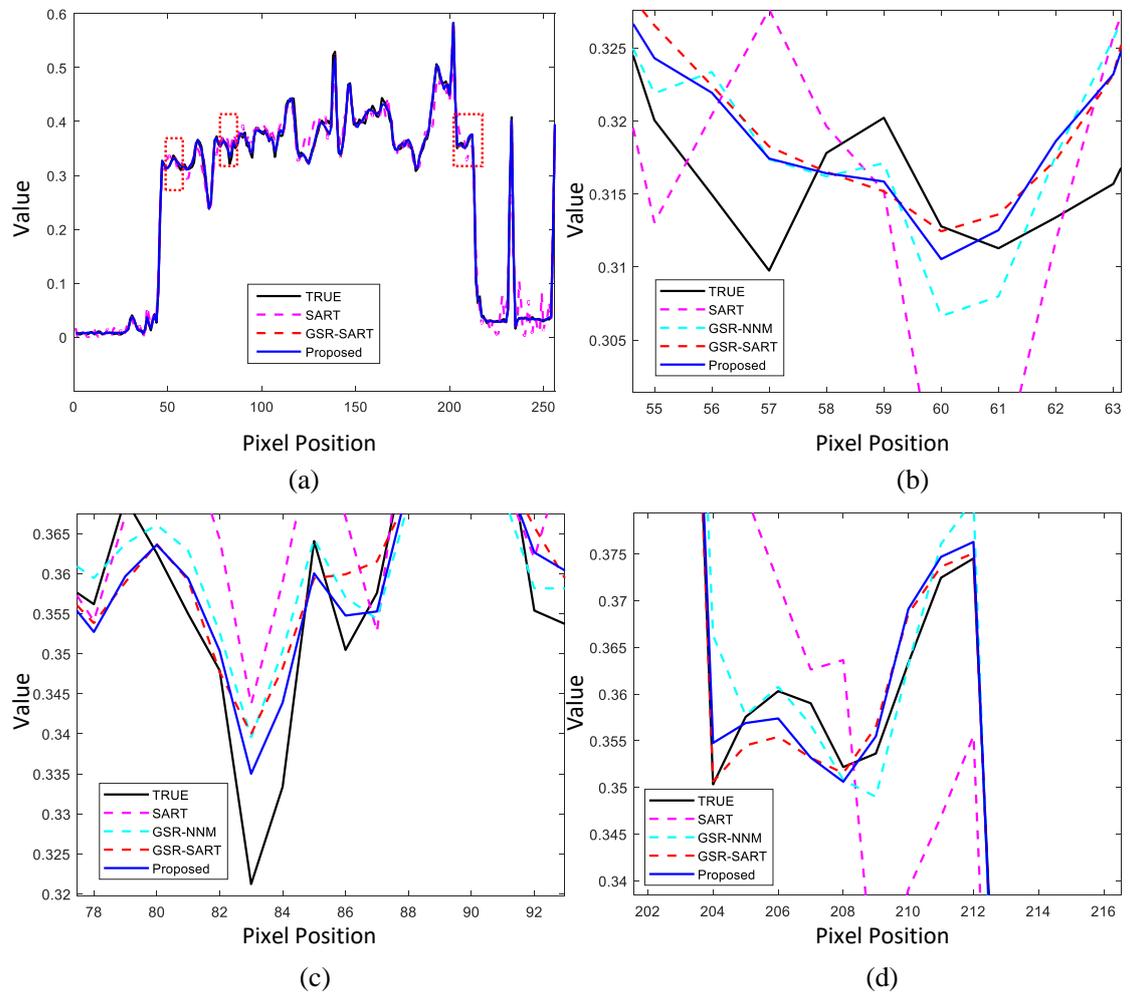

Fig. 4 Vertical profiles (128th column) of the abdominal and several generated profiles by different algorithms. (a) The overall profiles; (b)-(d) The magnified regions of our interest of CT image.

## 4.2. Thoracic

For the thoracic clinical data, the parameters are manually setting as $\lambda = 0.00008$ and $\mu = 0.15$. Table II shows all results achieved by different algorithms, which also shows that our proposed algorithm can achieve best quantitative results for all metric. The corresponding CT image results are shown in Fig. 5. In Fig. 5 (b), the results of SART still contains serious artifacts with its worst-case performance. In Fig. 5 (c), the convex algorithm of GSR-NNM may reconstruct CT image, but some artifacts may also exist. In Fig. 5 (d), the algorithm of GSR-SART may effectively reconstruct CT image but over-smoothed. Fig. 5 (e) shows that our proposed nonconvex method can achieve the best CT reconstruction result with more abundant details. Fig.5 (f)-(j) are corresponding magnified region of our interest of CT image (a)-(e), and Fig.5 (k)-(n) are reconstruction relative error abs($\mathbf{x} - \hat{\mathbf{x}}$) achieved by SART, GSR-NNM, GSR-SART and our proposed algorithm. All performances strongly demonstrate the ability of our proposed nonconvex $L_{1/2}$-regularized NSS denoiser, which can effectively preserve more structure and reduce artifacts for CT image reconstruction. Fig. 6 and Fig. 7 visualize the horizontal (128th row) and vertical (128th column) profiles of the abdominal image, respectively. In which, the profiles achieved by our proposed nonconvex algorithm are also obviously closer to the reference profiles.

**Table II**. Quantitative results reconstructed by different methods for thoracic

|  | PSNR | RMSE | SSIM |
|---|---|---|---|
| SART | 32.95 dB | 0.0225 | 0.8605 |
| GSR-NNM | 42.55 dB | 0.0075 | 0.9819 |
| GSR-SART | 43.27 dB | 0.0069 | 0.9803 |
| Proposed | **44.33 dB** | **0.0061** | **0.9844** |

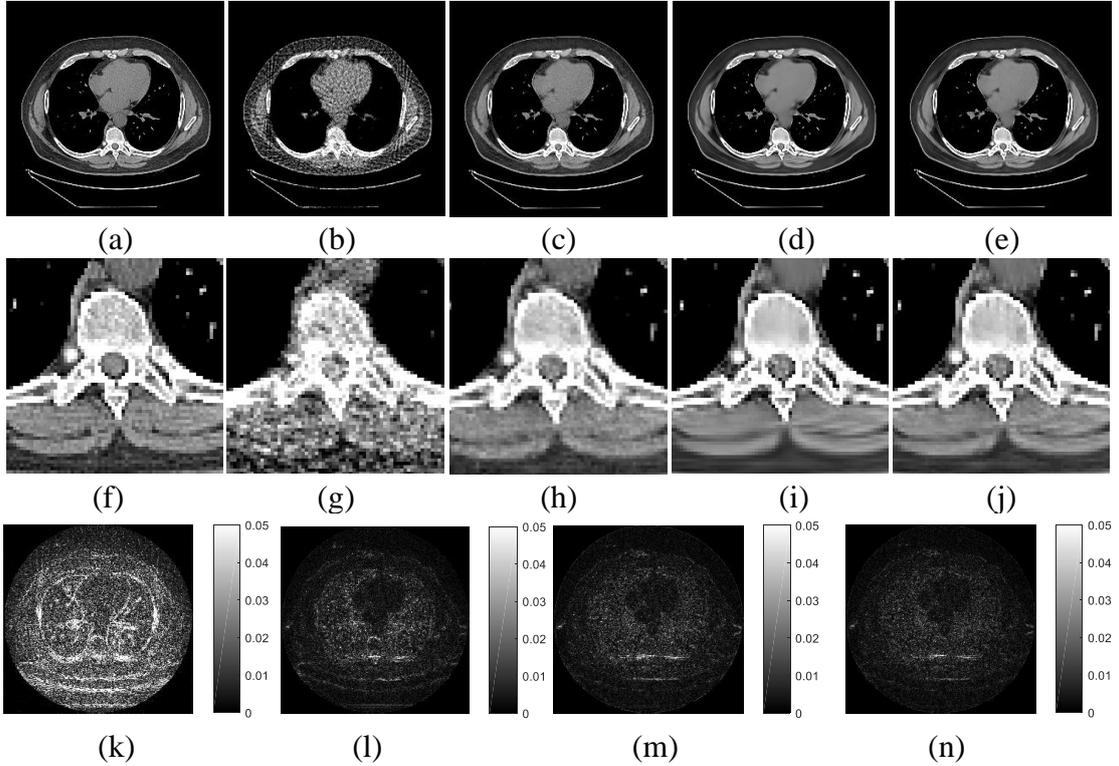

(a) (b) (c) (d) (e)

(f) (g) (h) (i) (j)

(k) (l) (m) (n)

Fig. 5 Reconstructed results of thoracic images. (a) Original image, (b) Reconstructed image by SART, (c) Reconstructed image by GSR-NNM, (d) Reconstructed image by GSR-SART, (e) Reconstructed image by our proposed nonconvex algorithm；(f)-(j) The magnified region of our interest of CT image (a)-(e), respectively, (k)-(n) The relative error achieved by SART, GSR-NNM, GSR-SART and our proposed algorithm.

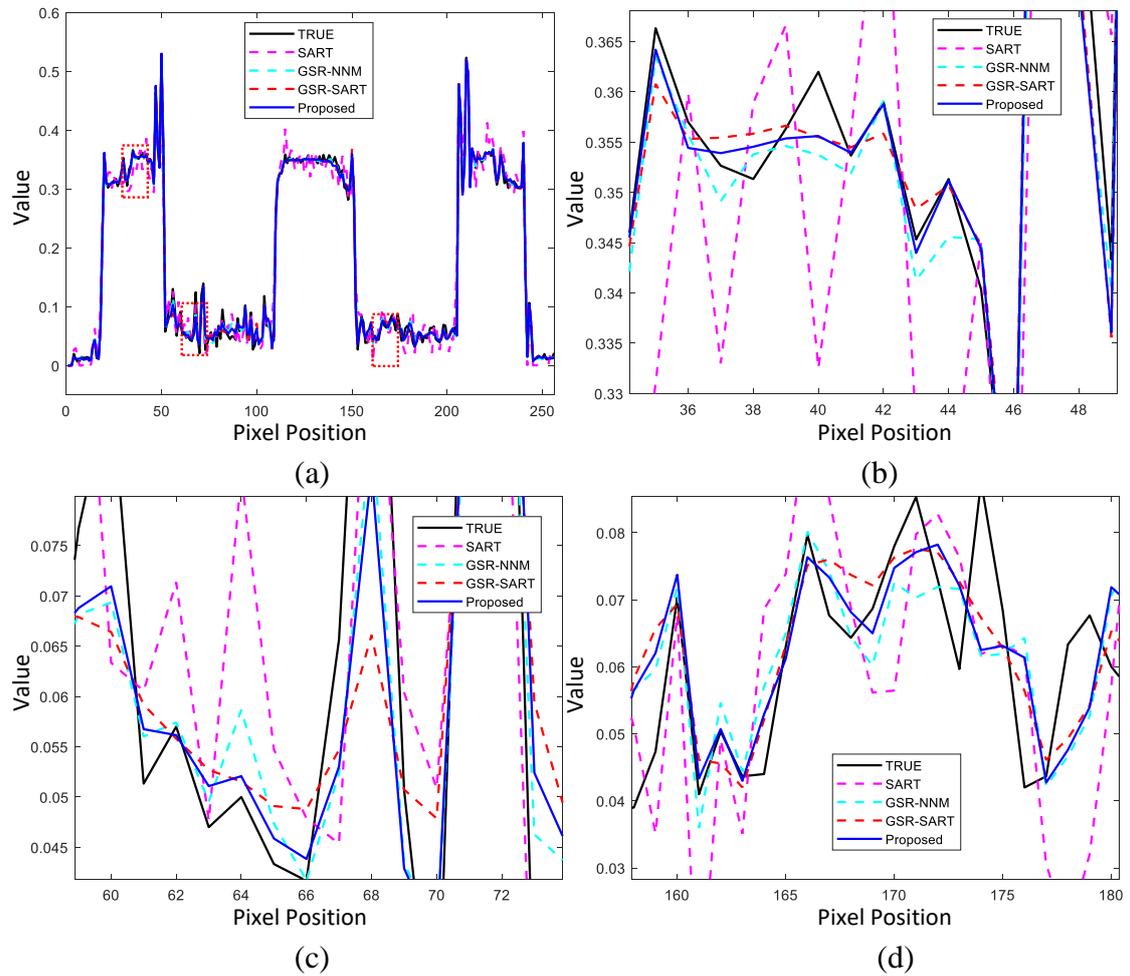

Fig. 6 Horizontal profiles (128th row) of the thoracic and several generated profiles by different algorithms. (a) The overall profiles; (b)-(d) The magnified regions of our interest of CT image.

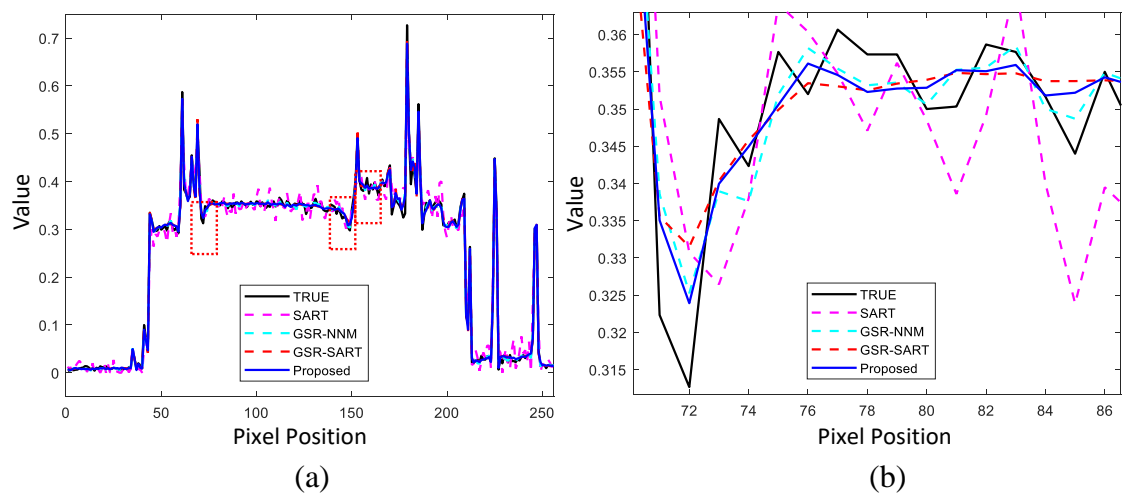

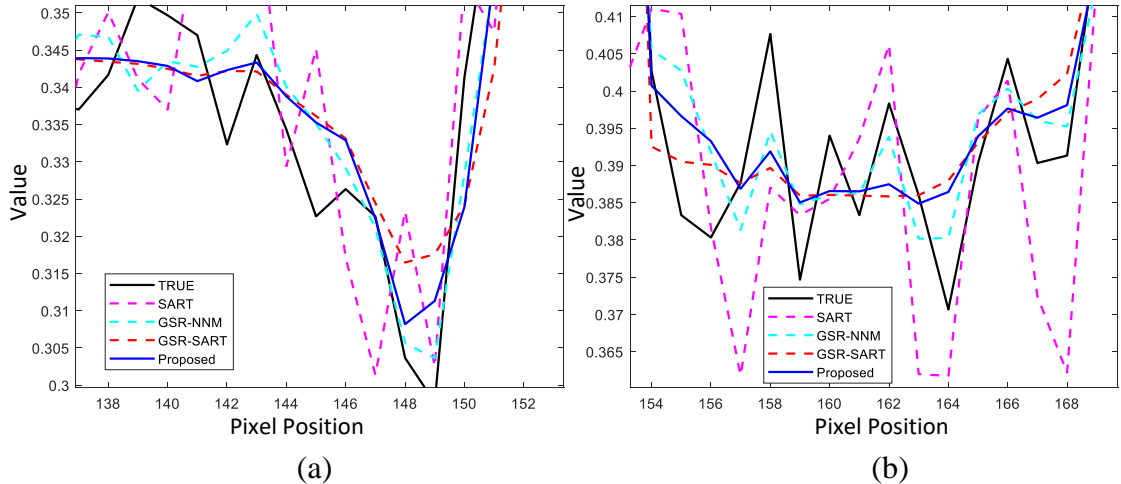

(a)                      (b)

Fig. 7 Vertical profiles (128th column) (128th row) of the thoracic and several generated profiles by different algorithms. (a) The overall profiles; (b)-(d) The magnified regions of our interest of CT image.

### 4.2. Pelvic

For the pelvic clinical data, the parameters are manually setting as $\lambda = 0.000025$ and $\mu = 0.16$. Table III shows all results achieved by different algorithms, which also shows that our proposed algorithm can achieve best quantitative results for all metric. The corresponding CT image results are shown in Fig. 8. Fig. 8 (b) presents the results achieved by SART, which also shows the worst-case performance. Fig. 8 (c) is the result of GSR-NNM with some artifacts may also exist. Fig. 8 (d) is obtained by the approach of GSR-SART and Fig. 8 (e) denotes the result by our proposed nonconvex method. Fig.8 (f)-(j) are corresponding magnified region of our interest of CT image (a)-(e). All performances strongly demonstrate the ability of our proposed nonconvex $L_{1/2}$ -group constraint denoiser, which can effectively preserve more structure and reduce artifacts for CT image reconstruction.

**Table III**. Quantitative results reconstructed by different methods for pelvic

|          | PSNR     | RMSE    | SSIM   |
|----------|----------|---------|--------|
| SART     | 33.29 dB | 0.0217  | 0.8772 |
| GSR-NNM  | 44.04 dB | 0.0063  | 0.9851 |
| GSR-SART | 46.30 dB | 0.0048  | 0.9912 |
| Proposed | **47.83 dB** | **0.0041** | **0.9902** |

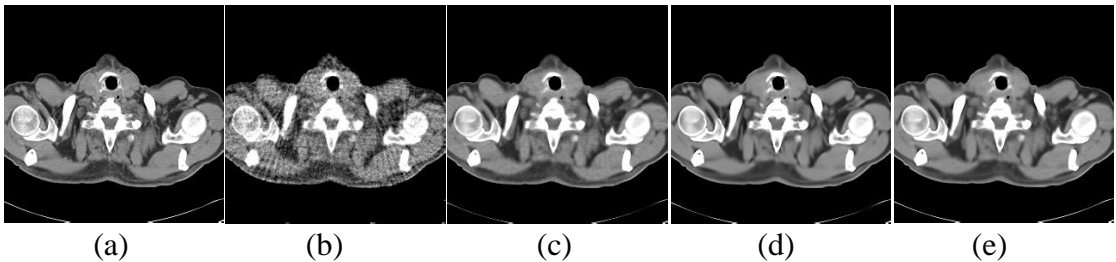

(a)       (b)       (c)       (d)       (e)

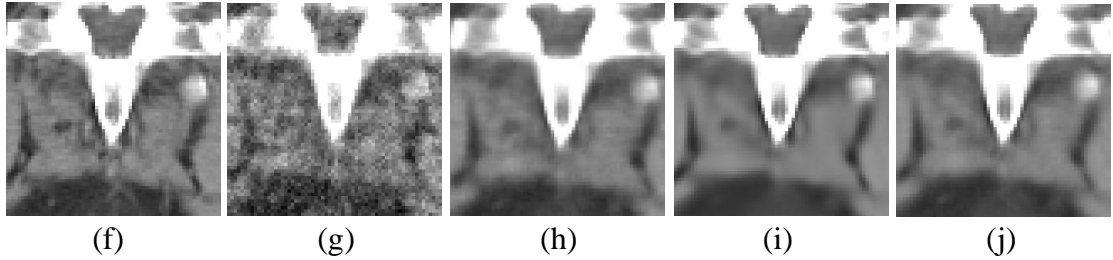

(f) (g) (h) (i) (j)

Fig. 8. Reconstructed results of pelvic images. (a) Original image, (b) Reconstructed image by SART, (c) Reconstructed image by GSR-NNM, (d) Reconstructed image by GSR-SART, (e) Reconstructed image by our proposed nonconvex algorithm；(f)-(j) The magnified region of our interest of CT image (a)-(e), respectively.

## V. Conclusion

A nonconvex $L_{1/2}$-norm regularized NSS denoiser model has been proposed for compressed sensing based CT reconstruction. Specifically, our CT reconstruction approach is developed by an alternating minimization approach, including SART and enhancement modules. To equivalently convert group sparse coding problem into low-rank approximation problem, we developed two different schemes by an adaptive dictionary learning strategy. Furthermore, we develop a fast iterative algorithm for nonconvex $L_{1/2}$ minimization problem by weighted singular value thresholding (WSVT) operator. All experimental results on three clinical CT images have demonstrated that our proposed NSS denoiser model can achieve better performance than popular approaches.

## Acknowledgments

The authors appreciate the anonymous reviewers for their extensive and informative comments for the improvement of this manuscript. This work was supported in part by the Educational Commission of Hunan Province of China under grant No: 21B0466.